\documentclass[prd,twocolumn,floatfix,english,letterpaper]{revtex4}

\usepackage[dvips]{graphicx}
\usepackage[T1]{fontenc}
\usepackage[latin9]{inputenc}
\usepackage{units}
\usepackage{amssymb}
\usepackage{esint}
\usepackage{babel}

\begin{document}

\title{Abelian Higgs model with charge conjugate boundary conditions}

\author{R.M. Woloshyn}
\affiliation{TRIUMF, 4004 Wesbrook Mall, Vancouver,
British Columbia, Canada V6T 2A3}

\begin{abstract}
The abelian Higgs model is studied on the lattice with charge conjugate
boundary conditions. A locally gauge invariant operator for the charged
scalar field is constructed and the charged scalar particle mass is
calculated in the Coulomb phase of the lattice model. Agreement
is found with the mass calculated in Coulomb gauge. The gauge invariant
scalar field operator is used to calculate the Higgs boson mass in
the Higgs region and to show that the charged particle disappears
from the spectrum in the confined regime. 
\end{abstract}

\maketitle

\section{Introduction}

The study of quantum chromodynamics using lattice field theory methods
has progressed to the stage where small effects due to electrodynamics
have to be considered. However, the description of charged particles
on a finite lattice with periodic boundary conditions which are typically
used in lattice simulations poses some challenges due to violation
of Gauss's law and the gauge dependence of the charged particle propagator.
See Refs. \cite{Gockeler:1991bu,Duncan:1996xy} for early work and 
\cite{Tantalo:2013maa,Portelli:2015wna} for reviews of recent developments.

Recently, Lucini {\it{et al}}. \cite{Lucini:2015hfa} have reconsidered the idea of using charge
conjugate boundary conditions \cite{Kronfeld:1990qu,Kronfeld:1992ae}. 
In this setup fields at positions differing
by a distance equal to the lattice size are related by charge conjugation.
In this way a charged particle on the lattice can have oppositely
charged images in neighboring lattice volumes and Gauss's law, which
is an obstruction in the case of periodic boundary conditions, can
be met. Furthermore, a gauge invariant form for the charged field
can be obtained.

In Ref. \cite{Lucini:2015hfa} the formalism for charged fields in a finite volume with charge
conjugate boundary conditions is set out and specific examples of
operators for lattice QED are constructed. In this work we apply the
ideas discussed in \cite{Lucini:2015hfa} to a theory of 
electrodynamics with scalar fields, namely,
the abelian Higgs model \cite{Higgs:1966ev}. 
The primary purpose is to illustrate the
calculation of the charged particle mass in a consistent gauge invariant
way in the Coulomb phase of the lattice model. In addition,
using a gauge invariant definition of the charged scalar field the
Higgs phenomenon and confinement, which are features of the lattice
Higgs model in other regions of the phase diagram 
\cite{Fradkin:1978dv,Jansen:1985cq,Jansen:1985nh}
, are demonstrated in a new way.

The general formalism for scalar field electrodynamics with charge conjugate
boundary conditions follows the development of Ref. \cite{Lucini:2015hfa}
and is given in Sec.~II. The specific lattice model used in this work is given
in Sec.~III. The results of lattice simulations are presented in Sec.~IV
In Sec.~\ref{sec:infin} the scalar model results in the absence of a gauge
field for periodic and charge conjugate boundary conditions are compared
to show that physics is not affected by boundary conditions. In subsequent
subsections of Sec.~IV some properties of the model in the Coulomb,
Higgs and confined regions are discussed. Sec.~V gives a summary.

\section{Formalism}

\subsection{General}

Consider the Euclidean space action for a complex scalar field $\phi$
with electrodynamics $S=S_{G}+S_{\phi}$ where 
\begin{equation}
\label{eq:gauge1}
S_{G}=\frac{1}{4}\int d^{4}x\, F_{\mu\nu}F\mu\nu
\end{equation}
 and 
\begin{eqnarray}
\label{eq:sca2}
\nonumber
S_{\phi} & = & \int d^{4}x [(D_{\mu}\phi(x))^{*}D_{\mu}\phi(x)+m_{c}^{2}\phi^{*}(x)\phi(x)\\
 &  & +\lambda_{c}(\phi^{*}(x)\phi(x))^{2}]
\end{eqnarray}
with $F_{\mu\nu}=\partial_{\mu}A_{\nu}(x)-\partial_{\nu}A_{\mu}(x)$
and $D_{\mu}=\partial_{\mu}+iqA_{\mu}(x).$ The action is invariant under
the transformations
\begin{equation}
\label{eq:Agt}
A_{\mu}(x)\rightarrow A_{\mu}(x)-\partial_{\mu}\alpha(x),
\end{equation}
and
\begin{equation}
\label{eq:phigt}
\phi(x)\rightarrow e^{iq\alpha(x)}\phi(x).
\end{equation}
We consider the theory in a finite cubic spatial volume with length
L on a side. The commonly used boundary conditions are periodic 
\begin{eqnarray}
A_{\mu}(x+L\hat{i}) & = & A_{\mu}(x),\\
\phi(x+L\hat{i}) & = & \phi(x)
\end{eqnarray}
for a shift $L$ in the $i$th direction. However, as discussed in
\cite{Lucini:2015hfa}, it is advantageous to charge conjugate when 
making a shift, that is, to apply the conditions
\begin{eqnarray}
A_{\mu}(x+L\hat{i}) & = & -A_{\mu}(x),\\
\phi(x+L\hat{i}) & = & \phi^{*}(x).
\end{eqnarray}
These charge conjugate boundary conditions are referred to as $C^{*}$
boundary conditions in \cite{Lucini:2015hfa}. In order to preserve the charge conjugate
boundary conditions the gauge transformation must also have a particular
form. Equation (3) implies that 
\begin{equation}
\label{eq:alpha}
\alpha(x)=\beta(x)+\textrm{constant},\;\beta(x+L\hat{i})=-\beta(x).
\end{equation}
 Then (\ref{eq:phigt}) requires that the constant in (\ref{eq:alpha}) should be an integer multiple
of ${\pi}/{q}.$ The most general gauge transformation is
therefore a combination of a local spatially anti-periodic function
and a global factor $\pm1$ acting on the scalar field. The global
phase symmetry of the action is broken by the boundary conditions from
$U(1)$ to $\mathbb{\mathbb{Z}}_{2}.$ 

The charge conjugate boundary conditions also affect the construction
of momentum eigenstates in finite volume \cite{Polley:1990tf,Lucini:2015hfa}. 
This has implications for
the lattice simulations that we carry out. The real part of the scalar
field is periodic and a zero momentum field can be constructed
by integrating over spatial positions. The mass of the particle associated
with the field can then be extracted directly from correlation function
of the projected field operator. On the other hand, the imaginary
part of the field is antiperiodic and in the lowest momentum eigenstate
there is a half unit of momentum ${\pi}/{L}$ associated to
each anti-periodic spatial direction. In the lattice simulation correlators
of real and imaginary fields have to be treated separately. The correlation
function of the imaginary part of the field yields an energy which
can be used in a dispersion relation to determine the mass.

\subsection{Charged field operator}

The construction of the charge field operator follows Ref. \cite{Lucini:2015hfa}. 
The charge $q$ may be a multiple of some elementary 
charge $Q={q}/{q_{el}}.$ Consider the operator 
\begin{equation}
\label{eq:genCCop}
\Phi_{J}(x)=e^{-iq\int d^{4}y\, A_{\mu}(y)J_{\mu}(y-x)}\phi(x)
\end{equation}
where $J_{\mu}(x)$ satisfies $\partial_{\mu}J_{\mu}(x)=\delta^{4}(x)$
and $J_{\mu}(x+L\hat{i})=-J_{\mu}(x).$ Note that a sign is changed 
compared to Eq. (3.1) in \cite{Lucini:2015hfa} to be consistent 
with the gauge transformation (\ref{eq:Agt}). Under a global transformation
$A_{\mu}(x)$ is invariant but 
\begin{equation}
\phi(x)\rightarrow e^{iq\alpha(x)}\phi(x)=(-1)^{Q}\phi(x)
\end{equation}
so 
\begin{equation}
\Phi_{J}(x)\rightarrow(-1)^{Q}\Phi_{J}(x).
\end{equation}
Using the properties of $J_{\mu}(x)$ and Eq. (\ref{eq:phigt}) it is easy to verify
that $\Phi_{J}(x)$ is invariant under a local (anti-periodic) gauge
transformation. 

Lucini {\it{et al}}. \cite{Lucini:2015hfa} give specific examples of 
functions $J_{\mu}(x)$ which
yield operators that can be used in a calculation. We adopt two of
them for this work. First consider a solution for $J_{\mu}(x)$ which
has the form
\begin{equation}
J_{0}(x)=0,\; J_{i}(x)=\delta(x_{0})\partial_{i}\Gamma({\mathbf x})
\end{equation}
where $\Gamma({{\mathbf x})}$ is anti-periodic.
Lucini {\it{et al}}. \cite{Lucini:2015hfa} give an explicit representation for $\Gamma$
but we do not need it here. Then the operator (\ref{eq:genCCop}) takes the form 
\begin{eqnarray}
\nonumber
\Phi_{J}(x) & = & e^{-iq\int d^{3}y\, A_{\mu}(x_0,{\mathbf y})\partial_{\mu}\Gamma({\mathbf y}-{\mathbf x})}\phi(x),\\
 & = & e^{iq\int d^{3}y\,\partial_{i}A_{i}(x_0,{\mathbf y})\Gamma({\mathbf y}-{\mathbf x})}\phi(x).
\end{eqnarray}
In Coulomb gauge $\partial_{i}A_{i}({\normalcolor {\normalcolor x})}=$0,
$\Phi_{J}(x)$ just becomes the gauge fixed scalar field which we
will denoted as $\phi_{c}(x).$ The correlator of the scalar field
in Coulomb gauge yields the gauge invariant mass for the charged particle.

Another solution is 
\begin{equation}
J_{\mu}(x)=\frac{1}{2}\delta_{\mu,i}\textrm{sgn}(x_{i})\prod_{\nu\neq i}\delta(x_{\nu}).
\end{equation}
The operator (\ref{eq:genCCop}) with this choice of $J_{\mu}(x)$, denoted as $\phi_{s}(x),$
takes the form 
\begin{equation}
\label{eq:ginvop}
\phi_{s}(x)=e^{i\frac{q}{2}\int_{-x_{i}}^{0}dsA_{i}(x+s\hat{i})}\phi(x)e^{-i\frac{q}{2}\int_{0}^{L-x_{i}}dsA_{i}(x+s\hat{i})}.
\end{equation}
The operator $\phi_{s}(x)$ consists of the scalar field with strings
emanating in the positive and negative $i$th spatial directions.
The strings join at the boundary and due to the boundary conditions
the operator is invariant under local gauge transformations. This
operator is a very convenient one for calculation since it can be
constructed easily without gauge fixing.

\section{Lattice formulation }

The lattice version of the abelian Higgs model has been extensively
studied, for example, in the pioneering work of Refs. 
\cite{Jansen:1985cq,Jansen:1985nh,Evertz:1986nt,Evertz:1986ur}.
With the compact
form of the lattice gauge field in terms of links $U_{x,\mu}=e^{iqA_{\mu}(x)}$
the lattice action $S=S_{G}+S_{\varphi}$ takes the form
\begin{eqnarray}
\label{eq:latpbc}
\nonumber
S_{G} & = & -\frac{\beta}{2}\sum_{P}(U_{P}+U_{P}^{*}),\\
S_{\varphi} & = & -\kappa\sum_{x,\mu}(\varphi^{*}(x)U_{x,\mu}\varphi(x+\hat{\mu)}+h.c.)\\
 &  & +\sum_{x}\varphi^{*}(x)\varphi(x)+\lambda\sum_{x}(\varphi^{*}(x)\varphi(x)-1)^{2}\nonumber
\end{eqnarray}
where $U_{P}$ is the product of links around the elementary plaquettes
and $\beta={1}/{q^{2}}$. This action is usually used with
periodic boundary conditions in all directions. The lattice field
and parameters are related to the continuum quantities in (\ref{eq:sca2}) by

\begin{eqnarray}
\phi(x) & = & \varphi(x)\sqrt{\kappa},\;\lambda_{c}=\frac{\lambda}{\kappa^{2}},\; m_{c}^{2}=\frac{1-2\lambda-8\kappa}{\kappa}.
\end{eqnarray}

With charge conjugate boundary conditions one would like to use the
gauge invariant operator (\ref{eq:ginvop}). As discussed in \cite{Lucini:2015hfa}
this is facilitated by introducing a lattice action where the matter field 
carries two units of charge. Following \cite{Lucini:2015hfa} the 
scalar QED version of the action is 
\begin{eqnarray}
\label{eq:latcc}
\nonumber
S_{G} & = & -2\beta\sum_{P}(U_{P}+U_{P}^{*}),\\
S_{\varphi} & = & -\kappa\sum_{x,\mu}(\varphi^{*}(x)(U_{x,\mu})^{2}\varphi(x+\hat{\mu)}+h.c.)\\
 &  & +\sum_{x}\varphi^{*}(x)\varphi(x)+\lambda\sum_{x}(\varphi^{*}(x)\varphi(x)-1)^{2}\nonumber
\end{eqnarray}
which will be implemented with charge conjugate boundary conditions
in all spatial directions and periodic in time. This action is invariant
under the local gauge transformations
\begin{eqnarray}
U_{x,u} & \rightarrow & \Lambda_{x}U_{x,\mu}\Lambda_{x+\hat{\mu}}^{*},\\
\varphi(x) & \rightarrow & \Lambda_{x}^{2}\varphi(x)
\end{eqnarray}
where the transformation $\Lambda$ satisfies $\Lambda_{x+L\hat{i}}=\Lambda_{x}^{*}.$ 

To investigate the properties of charged field the scalar field after
Coulomb gauge fixing $\varphi_{c}$ will be used as well as the lattice
version of (\ref{eq:ginvop}) which takes the form
\begin{equation}
\label{eq:opphis}
\varphi_{s}(x)=\prod_{s=-x_{i}}^{-1}U_{x+s\hat{i},i}\varphi(x)\prod_{s=0}^{L-x_{i}-1}U_{x+s\hat{i},i}^{*}.
\end{equation}

\section{Results}

\begin{figure}
\scalebox{0.50}{\includegraphics*{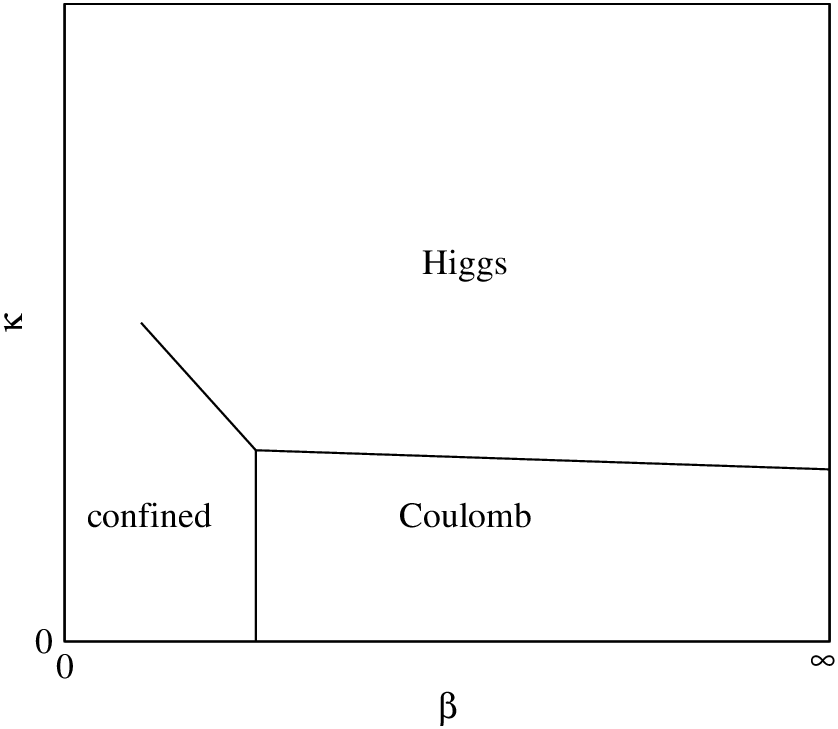}}
\caption{Schematic phase diagram for the lattice abelian Higgs model at 
fixed $\lambda$.}
\label{fig:fig1}
\end{figure}

The phase diagram for the lattice abelian Higgs model 
\cite{Jansen:1985cq,Jansen:1985nh} at a fixed $\lambda$
is shown schematically in Fig. \ref{fig:fig1}. One can identify three regions:
confined, Higgs and Coulomb. However, the confined
and Higgs regimes do not actually correspond to distinct phases as
they can be connected by analytic continuation around the transition
line that separates the confined and Higgs regions \cite{Fradkin:1978dv}. 
For $\lambda\gtrsim0.1$
the transition line ends at a value of $\beta$ greater than 0 as
shown in the figure. Free charges are expected to exist only in the
Coulomb phase \cite{Fradkin:1978dv}.

The lattice simulations presented here were carried out on $16^{4}$
site lattices using a multi-hit Metropolis updating algorithm. The
primary goal is to explore the calculation of the charged particle
mass in the Coulomb phase. Evertz {\it{et al}}. \cite{Evertz:1986nt}
studied charged particle
mass in the abelian Higgs model long ago using an indirect method.
To make some contact with this earlier work we choose the same values
$\beta$ = 2.5, $\lambda$ = 3 for most of the mass calculations. To 
illustrate the confining feature of the model some calculations at 
other values of $\beta$ were also done.

\subsection{\label{sec:infin}$\beta=\infty$}

In the absence of the gauge field, corresponding to $\beta=\infty$,
the model reduces to a $\varphi^{4}$ theory. As a preliminary step
we compare calculations with periodic and charge conjugate spatial
boundary conditions (periodic in time) at $\beta=\infty$. Ensembles
of 32,000 scalar field configurations were used in these calculations\@.

\begin{figure}
\scalebox{0.50}{\includegraphics*{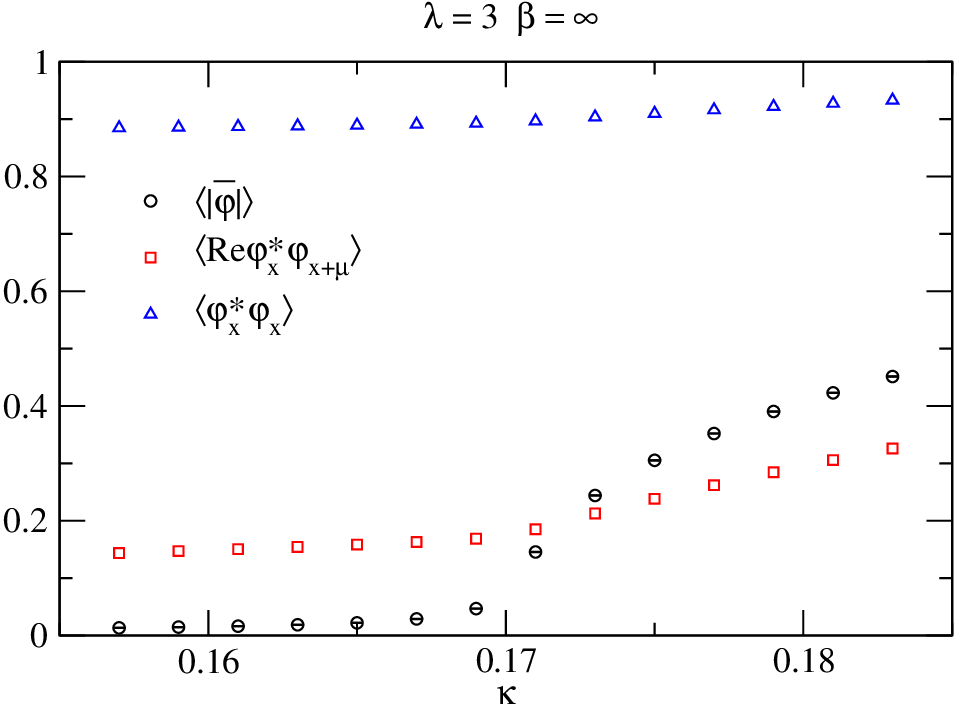}}
\caption{Observables calculated using the scalar field action $S_{\varphi}$ with periodic 
boundary conditions in the absence of a gauge field.}
\label{fig:fig2}
\end{figure}

\begin{figure}
\scalebox{0.50}{\includegraphics*{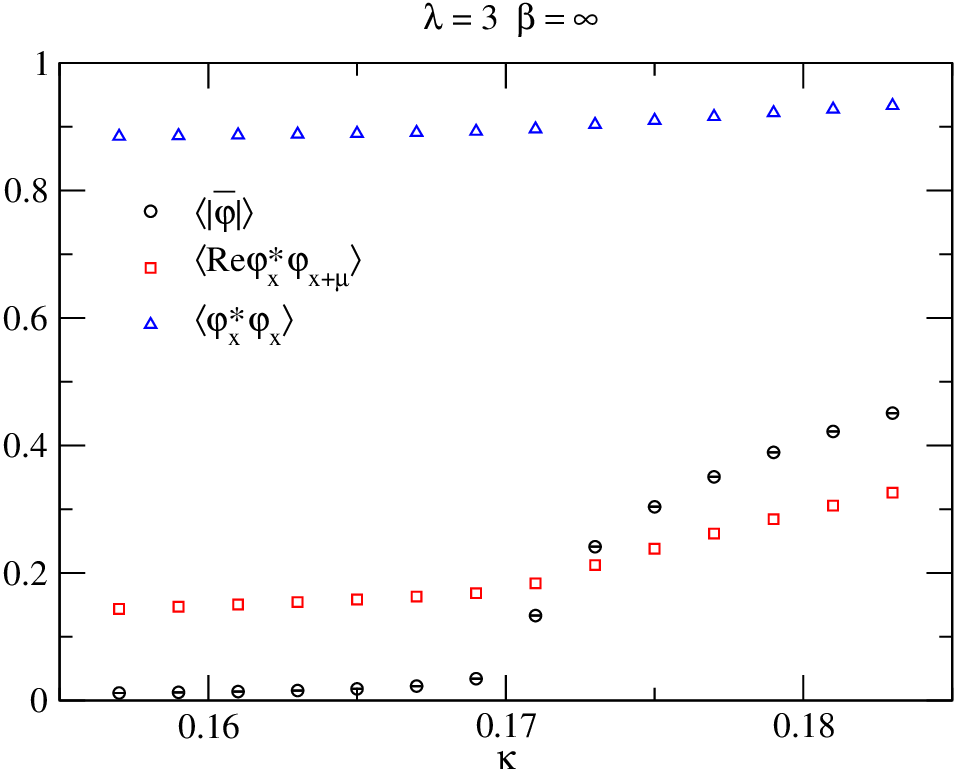}}
\caption{Observables calculated using the scalar field action $S_{\varphi}$ with charge 
conjugate boundary conditions in the absence of a gauge field.}
\label{fig:fig3}
\end{figure}

When $\kappa$ goes from small to large values there is a transition
to a spontaneously broken symmetry phase. To calculate the vacuum
expectation value of $\varphi,$ which would serve as an order parameter,
one should introduce a symmetry breaking term with an external field
$\eta$, for example, $\eta\varphi$ into the action, calculate $\left\langle \varphi\right\rangle $
and take the thermodynamic and $\eta\rightarrow0$ limits. However,
there is a simpler procedure without an external field which provides
a reasonable estimator for $\left\langle \varphi\right\rangle $(see
Ref. \cite{Hasenfratz:1988kr}). Consider the field averaged over a 
single configuration with lattice volume $V$
\begin{equation}
\overline{\varphi}=\frac{1}{V}\sum_{x}\varphi(x),
\end{equation}
 and the projection of $\varphi$ in the direction of $\overline{\varphi}$
\begin{equation}
\widetilde{\varphi}(x)=\frac{\varphi^{*}(x)\overline{\varphi}}{\left|\overline{\varphi}\right|}.
\end{equation}
Then the expectation value
\begin{equation}
\left\langle \widetilde{\varphi}\right\rangle =\left\langle \left|\overline{\varphi}\right|\right\rangle 
\end{equation}
will be used as proxy for $\left\langle \varphi\right\rangle .$
The results for $\left\langle \left|\overline{\varphi}\right|\right\rangle $
as a function of $\kappa$ at $\lambda$ = 3 are shown in Fig. \ref{fig:fig2} and
Fig. \ref{fig:fig3} for simulations with periodic and charge conjugate spatial
boundary conditions respectively. The transition in the vicinity of
$\kappa$ = 0.17 is seen clearly. The expectation values of the operators
Re$(\varphi^{*}(x)\varphi(x+\hat{\mu))}$ and $\varphi^{*}(x)\varphi(x)$
which appear in the action are also shown. Although these are not
strictly speaking order parameters their behavior as function of $\kappa$
can give an indication that the theory undergoes a transition. Simulations
with periodic and charge conjugate boundary conditions yield compatible
results on our $16^{4}$ lattice.

\begin{figure}
\scalebox{0.50}{\includegraphics*{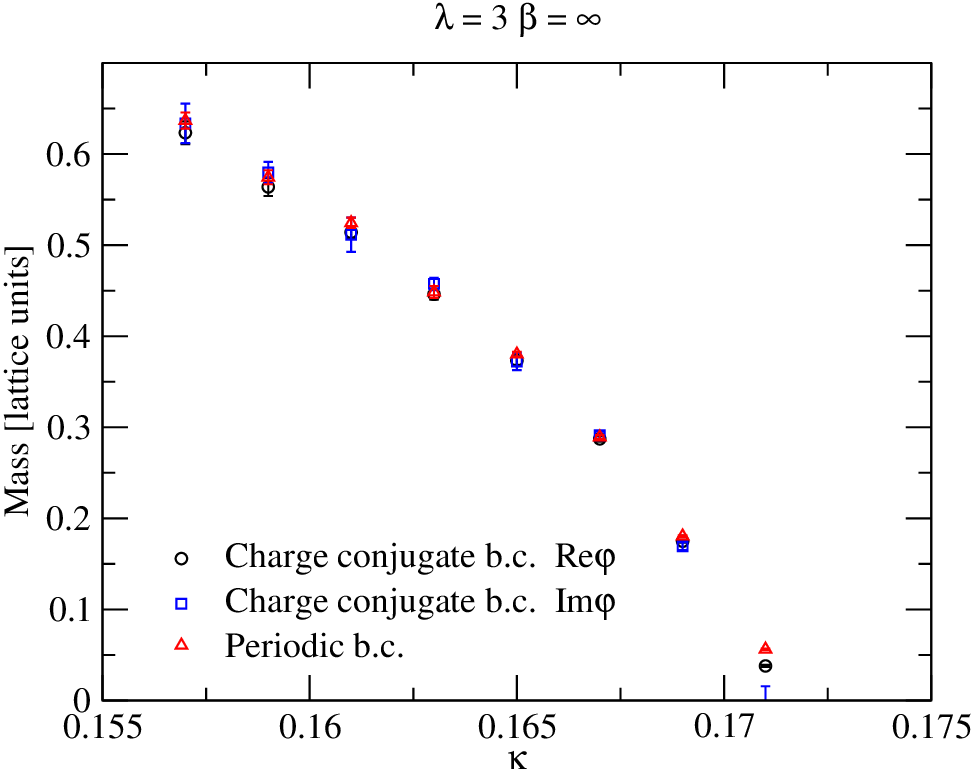}}
\caption{Scalar particle mass in lattice units as a function of $\kappa$
calculated using the scalar field action in the absence of a gauge field.}
\label{fig:fig4}
\end{figure}

Since the gauge field is absent correlation functions of $\varphi$
can be used directly to calculate the scalar mass. The results in
the symmetric phase are shown in Fig. \ref{fig:fig4}. Recall that with charge
conjugate boundary conditions Im$\varphi$ is anti-periodic in space
so Im$\varphi$ is projected to momentum $({\pi}/{L})$(1,1,1).
The energy extracted from the correlator of the momentum projected
Im$\varphi$ field is converted to a mass using the lattice dispersion
relation 
\begin{equation}
\label{eq:latdr}
2\cosh(E)=m^{2}+8-6\cos({\pi}/{L}).
\end{equation}
The mass determined this way is consistent with the mass extracted
from the zero-momentum correlator of Re$\varphi$ as it should be.

\subsection{Finite $\beta$ }

\begin{figure}
\scalebox{0.50}{\includegraphics*{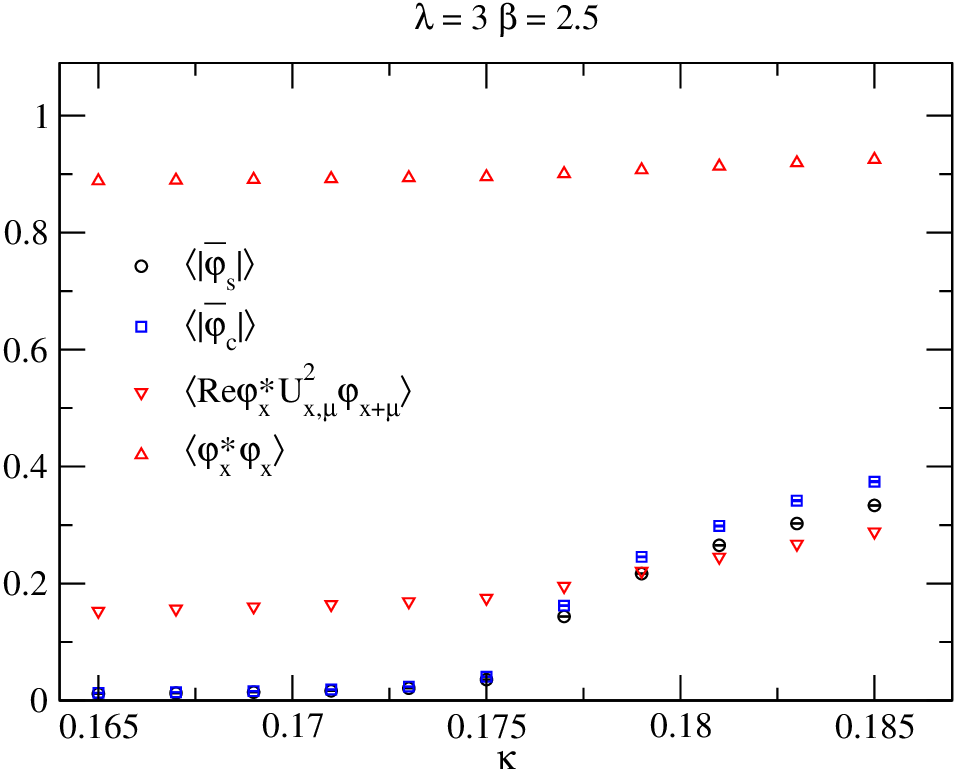}}
\caption{Observables as a function of $\kappa$ calculated with the action
(\ref{eq:latcc}) using charge conjugate boundary conditions.} 
\label{fig:fig5}
\end{figure}

Having seen that periodic and charge conjugate boundary conditions
give compatible results in the absence of a gauge field we turn in
this section to the model at finite $\beta$. Figure \ref{fig:fig5} shows observables
calculated at fixed values of $\beta$ and $\lambda$ (2.5 and 3 respectively)
as a function of $\kappa$ on a $16^{4}$ lattice with the action
(\ref{eq:latcc}) using charge conjugate boundary conditions. The expectation values
of $\left|\overline{\varphi_{s}}\right|$ and $\left|\overline{\varphi_{c}}\right|$
show clearly the transition from the Coulomb phase to the Higg
regime in the vicinity of $\kappa$ equal to 0.177. This is very close
to the transition point found in Ref. \cite{Evertz:1986ur}
using the action (\ref{eq:latpbc}) 
with periodic boundary conditions and with the same values 
of $\beta$ and $\lambda$.
This gives confidence that the physics of the lattice Higgs model
used in this work is the same as that of the action (\ref{eq:latpbc}) 
used in earlier work.

The primary objective here is to demonstrate the calculation of the
charged scalar boson mass in the Coulomb phase. This corresponds
to the region that would be relevant for the use of a lattice U(1)
gauge theory in more realistic applications such as electromagnetic
corrections to QCD. Correlation functions of four different scalar
field operators Re$\varphi_{s},$ Im$\varphi_{s},$ Re$\varphi_{c},$
Im$\varphi_{c}$ were analyzed. Recall that the imaginary parts of
the field are anti-periodic in spatial directions so for these fields
projection to momentum $({\pi}/{L})$(1,1,1) is carried out
and mass is determined from energy using (\ref{eq:latdr}).
The real components of the field are projected to zero 
momentum in the usual way.

\begin{figure}
\scalebox{0.50}{\includegraphics*{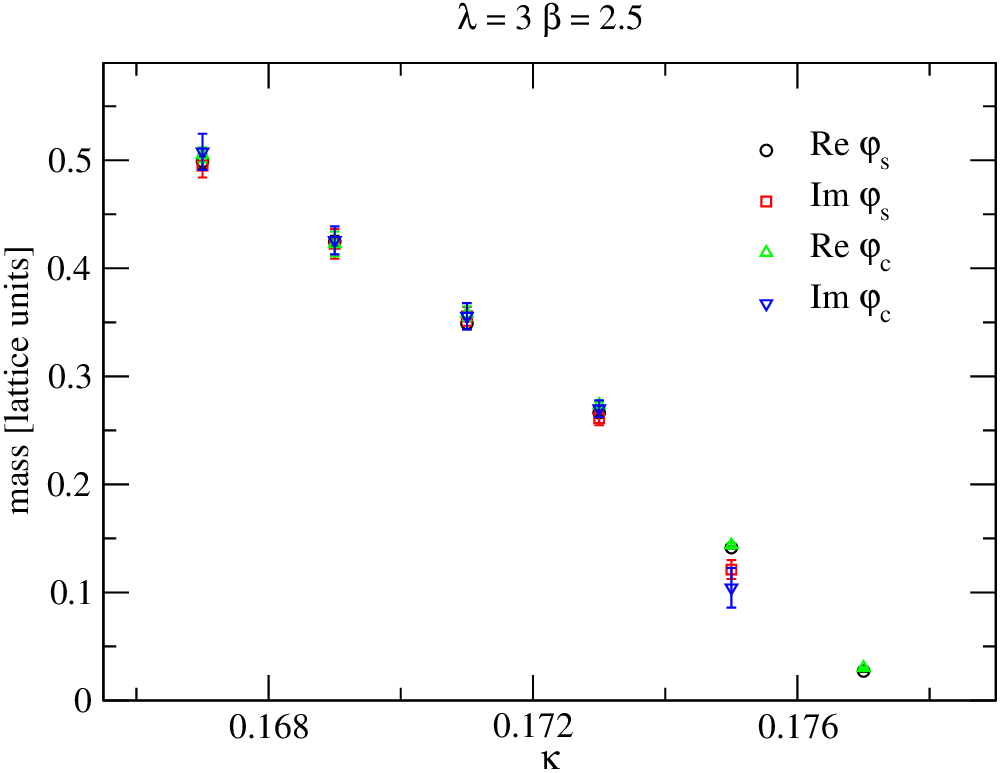}}
\caption{Charged particle mass in lattice units as a function of $\kappa$ 
in the Coulomb phase calculated with different operators.}
\label{fig:fig6}
\end{figure}

For calculations with the gauge invariant operator an ensemble of
32,000 field configurations was used. These were constructed with
a multi-hit Metropolis algorithm with 30 sweeps between saved configurations.
Since gauge fixing is rather time consuming the Coulomb gauge fixed
sample had only 8,000 configurations. The Euclidean time correlation
functions were fit with two exponential terms (symmetrized in time).
Statistical errors were calculated by a jackknife procedure. The masses
in lattice units are plotted in Fig. \ref{fig:fig6}. There is good 
consistency between
different determinations over a range of $\kappa$ values which provides some
confidence that the formulation of the lattice theory presented in 
\cite{Lucini:2015hfa} can be used effectively to deal with charged particles.

\subsection{Higgs phase}

\begin{figure}
\scalebox{0.50}{\includegraphics*{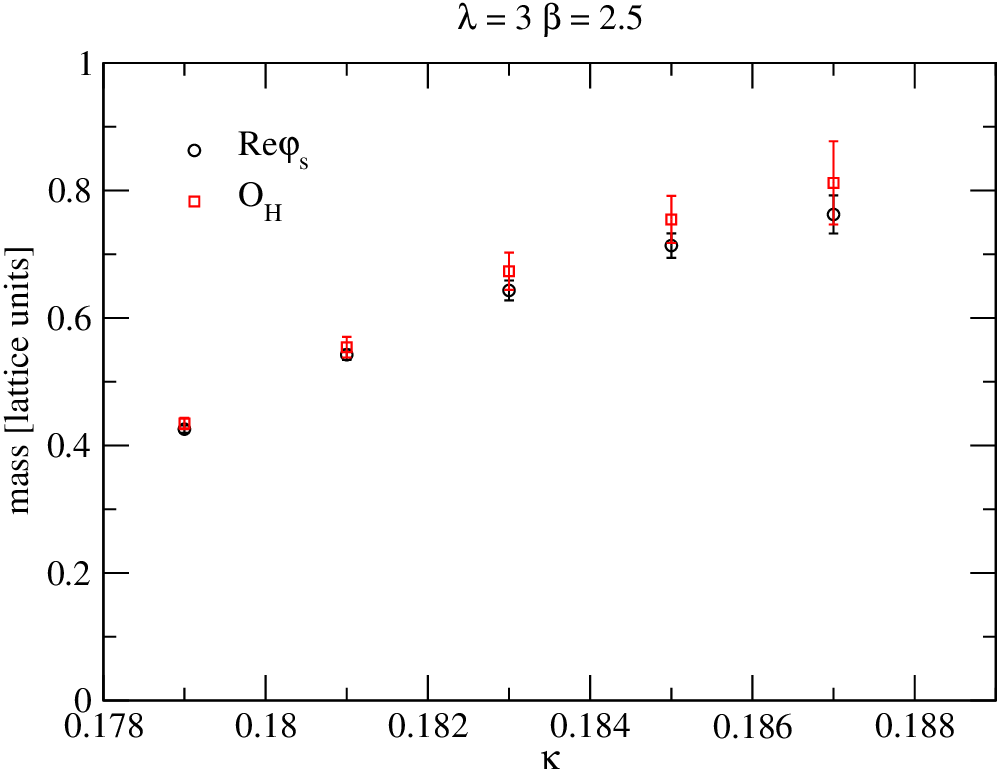}}
\caption{Scalar particle mass in lattice units as a function of $\kappa$ 
in the Higgs region calculated with different operators.}
\label{fig:fig7}
\end{figure}

Since the use of charge conjugate boundary conditions allows for a
gauge invariant operator for the scalar field we have a new way to
explore the Higgs region. In the standard semi-classical treatment
of the Higgs phenomenon the Higgs boson is an elementary field. 
In contrast, from the nonperturbative perspective of the lattice Higgs
model the Higgs boson has been interpolated using a gauge invariant
composite operator \cite{Evertz:1986ur}. For the action (\ref{eq:latcc}) 
the composite Higgs operator takes the form
\begin{equation}
\label{eq:ophiggs}
O_{H}=\textrm{Re}\sum_{i}\varphi^{*}(x)U_{x,i}^{2}\varphi(x+i)
\end{equation}
where the sum is over spatial directions. The construction (\ref{eq:opphis})
provides a locally gauge invariant scalar field and it is natural to ask if
it also describes the Higgs boson. Correlation functions of Re$\varphi_{s}$
and $O_{H}$ (with vacuum expectation values subtracted) were analyzed
in the Higgs region above $\kappa$ = 0.177. The mass in lattice units
is shown in Fig. \ref{fig:fig7}. The statistical errors are from a jackknife analysis.
The masses extracted using the two different fields are consistent
over the range of $\kappa$ values that were investigated. At the upper
end of this range the statistical uncertainties are growing so to
go to even larger $\kappa$ would require field ensembles much larger 
than those used in this study.

The field $\varphi_{s}$ is composite but in way that is different
from $O_{H}.$ It consists of the elementary field $\varphi$ with
a cloud of gauge field fluctuations. It gives a view of the Higgs
phenomenon which has some similarity to the semi-classical treatment
\cite{Higgs:1966ev}
but without the notion of spontaneous local gauge symmetry breaking
which, in the nonperturbative framework, would not be viable
\cite{Elitzur:1975im}.

\subsection{Confinement}

\begin{figure}
\scalebox{0.50}{\includegraphics*{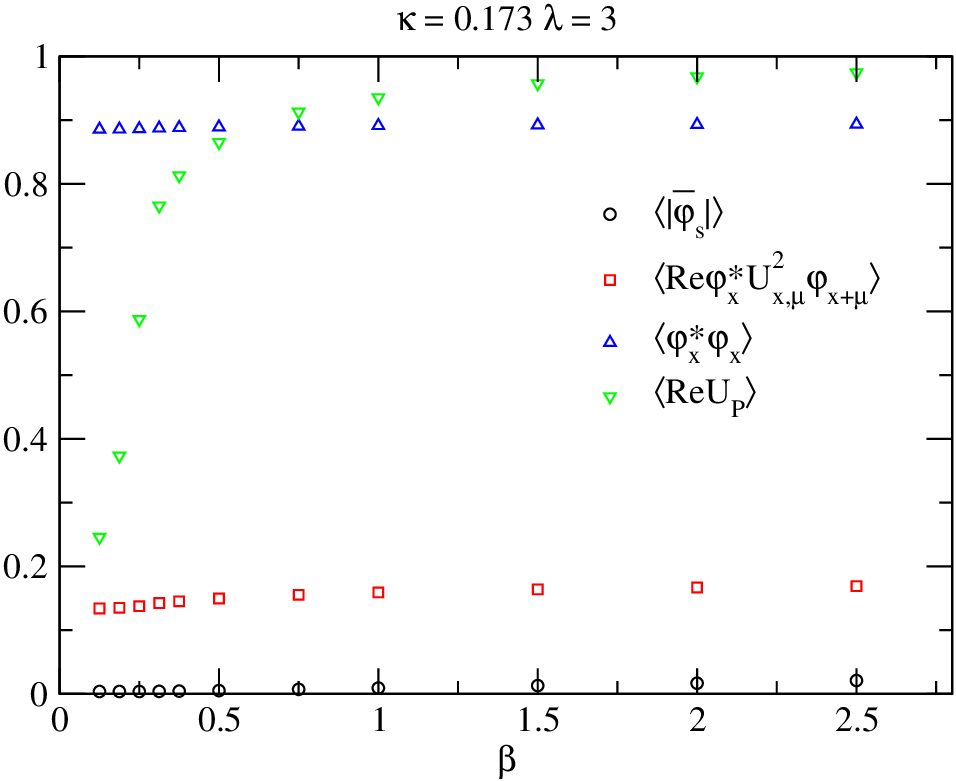}}
\caption{Observables as a function of $\beta$ calculated with the action
(\ref{eq:latcc}) with charge conjugate boundary conditions.}
\label{fig:fig8}
\end{figure}

\begin{figure}
\scalebox{0.50}{\includegraphics*{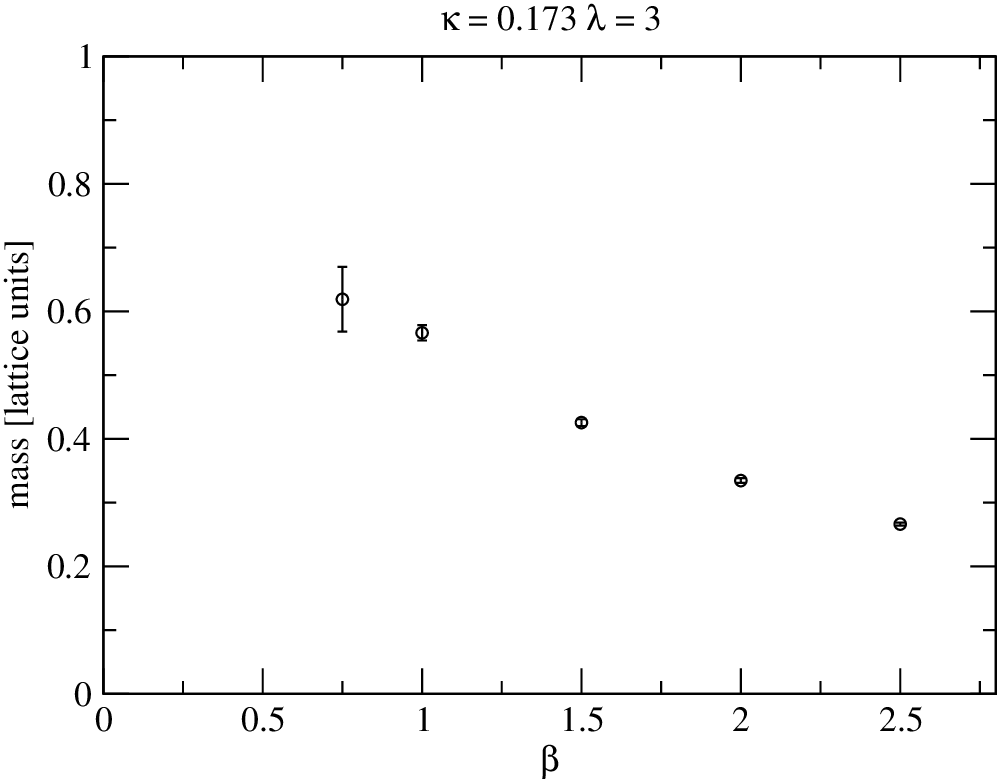}}
\caption{Charged particle mass in lattice units as a function of $\beta$ 
calculated using the operator Re$\varphi_{s}$.}
\label{fig:fig9}
\end{figure}

At small $\beta$ compact lattice QED is confining. We explore the
transition to the confined regime by calculating at fixed $\kappa$ and
$\lambda$ and decreasing $\beta$ starting a point in the Coulomb
phase. Figure \ref{fig:fig8} shows the values of some observables as a function
of $\beta$. The gauge field plaquette variable Re$U_{P}$ shows the
transition from the weak coupling to the strong coupling regime around
$\beta$ = 0.25. The vacuum expectation values of observables involving
the $\varphi$ field are quite insensitive to the value of $\beta$
and exhibit only small changes in the transition from weak coupling
to strong coupling. The mass of the charged scalar extracted from the
correlation function of Re$\varphi_{s}$ increases steadily as $\beta$
is decreased as shown in Fig. \ref{fig:fig9}. 
Below $\beta$ = 0.75 the correlation function falls
very rapidly as function of time so even with an ensemble of 32,000
configurations it was not possible to make an accurate mass determination.
\begin{figure}
\scalebox{0.50}{\includegraphics*{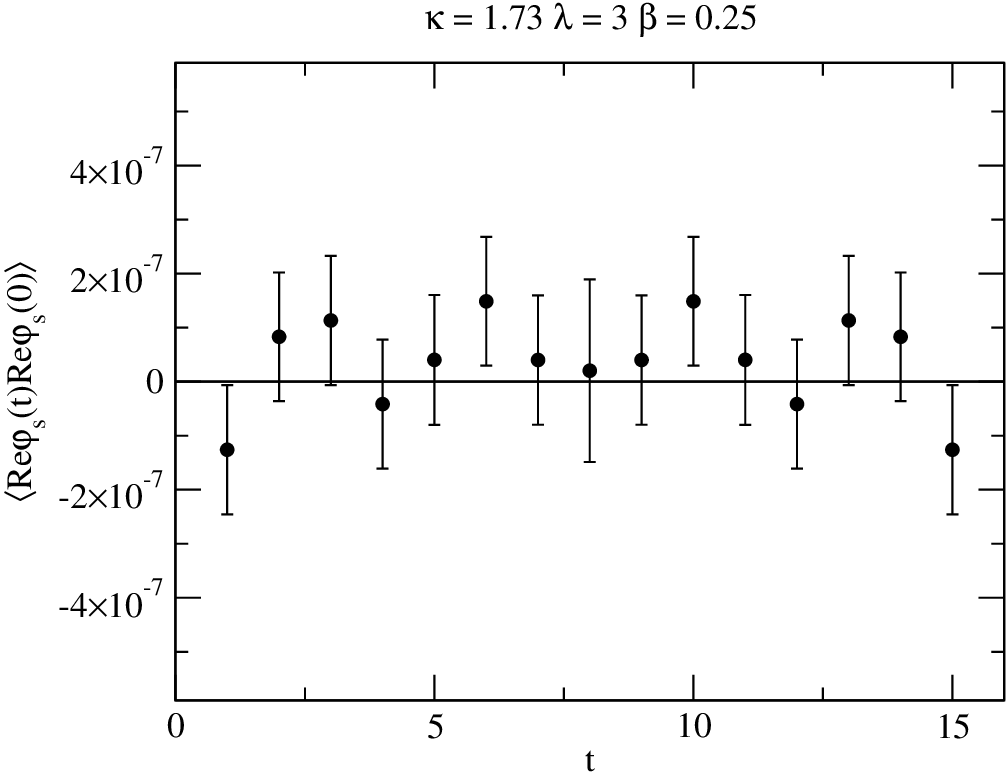}}
\caption{Correlation function 
$\left\langle\textrm{Re}\varphi_{s}(t)\textrm{Re}\varphi_{s}(0)\right\rangle$
at $\beta$ = 0.25.}
\label{fig:fig10}
\end{figure}
Figure \ref{fig:fig10} shows the correlation function 
$\left\langle\textrm{Re}\varphi_{s}(t)\textrm{Re}\varphi_{s}(0)\right\rangle$
at $\beta$ = 0.25. In this
region the correlation function is just noise. The charged scalar
field does not propagate. It has disappeared from the spectrum which
can be taken as a signature of confinement.

\begin{figure}
\scalebox{0.50}{\includegraphics*{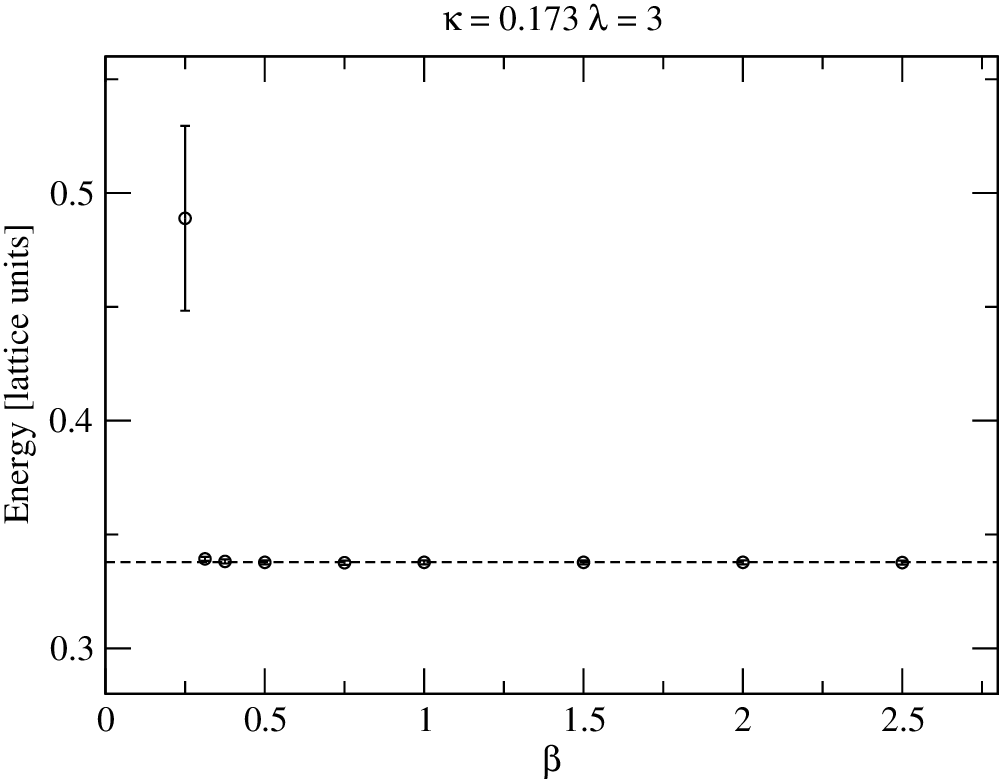}}
\caption{Ground state energy from the correlation function 
of the operator $O_p$ (Eq. (\ref{eq:photop})) as a
function of $\beta$ calculated at momentum $({\pi}/{L})$(1,1,1).
The dashed line shows the energy of a zero mass particle at this 
momentum from the dispersion relation (\ref{eq:latdr}).
}
\label{fig:fig11}
\end{figure}

In the strong coupling region the gauge field should also be confined.
This can be demonstrated using the photon propagator. For the photon
interpolating operator one can use
\begin{equation}
O_{p}=\textrm{Im}\sum_{i,j}U_{P}
\label{eq:photop}
\end{equation}
which is the imaginary part of the gauge field plaquette summed over
spatial planes \cite{Evertz:1986ur}. In the Coulomb phase the photon 
is expected
to be massless \cite{Fradkin:1978dv} so the correlation function should be
calculated at a nonzero momentum. We use momentum $({\pi}/{L})$(1,1,1)
consistent with our boundary conditions. The energy calculated from
the momentum projected correlation function of $O_{p}$ is plotted
in Fig. (\ref{fig:fig11}). The dashed line is the shows the energy for a zero mass
particle calculated using the dispersion relation (\ref{eq:latdr}).
In the Coulomb phase the gauge field correlator is consistent with 
the presence of a zero mass photon. Around $\beta$ = 0.25 the mass departs from 
zero and at smaller values of $\beta$ the correlator of $O_{p}$ is reduced
to noise similar to what is seen in Fig. \ref{fig:fig10} signaling the confinement
of the gauge field.

\section{Summary}

The use of charge conjugate boundary conditions, as discussed by Lucini
{\it{et al}}. \cite{Lucini:2015hfa}, provides an interesting option for 
dealing with QED on the lattice. An attractive feature of this formulation 
is that the mass of the charged field can be determined using a simple gauge 
invariant procedure. In this paper we have implemented the ideas of 
\cite{Lucini:2015hfa} in a lattice theory of electrodynamics with scalar
fields, the abelian Higgs model. 

In Sect. 4.1 the model in the absence of a gauge field ($\beta=\infty$)
is compared for charge conjugate and periodic boundary conditions.
The results for a variety of observables are compatible. At finite
$\beta$ and other parameters within the pertubative region of the
model the charged scalar mass was calculated using both gauge invariant
(Eq. (\ref{eq:opphis})) and Coulomb gauge fixed fields. Due to the choice of 
boundary conditions the imaginary parts of the fields require projection to
a non-zero momentum with mass determined using the lattice dispersion
relation (\ref{eq:latdr}). As shown in Fig. (\ref{fig:fig6}) these technically 
varied procedures yield compatible charged particle masses.

The gauge invariant field $\varphi_{s}$ is also useful for exploring
the Higgs model in other regions of the phase diagram. In the Higgs
regime the correlator of Re$\varphi_{s}$ gives masses which are compatible
with those extracted using the composite scalar operator (\ref{eq:ophiggs})
which has been used in the past to interpolate the Higgs boson. In the strong
coupling confining region we showed that the particle associated with
field $\varphi_{s}$ disappears from the physical spectrum.

In summary, this works demonstrates the efficacy of the formulation
of \cite{Lucini:2015hfa} for numerical studies of lattice U(1) gauge theory 
and encourages further applications.

\acknowledgments
It is a pleasure to thank C.~Itoi for a very helpful discussion.
TRIUMF receives federal funding via a contribution agreement 
with the National Research Council of Canada.



\begin{thebibliography}{99}

\bibitem{Gockeler:1991bu}
M.~G\"{o}ckeler, R.~Horsley, P.~Rakow, G.~Schierholz, and R.~Sommer, 
Nucl. Phys. \textbf{B371}, 713 (1992).

\bibitem{Duncan:1996xy}
A.~Duncan, E.~Eichten, and H.~Thacker,
Phys. Rev. Lett. \textbf{76}, 3894 (1996).

\bibitem{Tantalo:2013maa}
N.~Tantalo,
PoS \textbf{LATTICE2013}, 007 (2014).

\bibitem{Portelli:2015wna}
A.~Portelli,
PoS \textbf{LATTICE2014}, 013 (2015).

\bibitem{Lucini:2015hfa}
B.~Lucini, A.~Patella, A.~Ramos, and N.~Tantalo,
JHEP \textbf{1602}, 076 (2016).

\bibitem{Kronfeld:1990qu}
A.~S.~Kronfeld, and U.-J.~Wiese,
Nucl. Phys. \textbf{B357}, 521 (1991).

\bibitem{Kronfeld:1992ae}
A.S.~Kronfeld, and U.-J.~Wiese,
Nucl. Phys. \textbf{B401}, 190 (1993).

\bibitem{Higgs:1966ev}
P.W.~Higgs,
Phys. Rev. \textbf{145}, 1156 (1966).
 
\bibitem{Fradkin:1978dv}
E.H.~Fradkin and S.H.~Shenker,
Phys. Rev. D \textbf{19}, 3682 (1979).

\bibitem{Jansen:1985cq}
K.~Jansen, J.~Jers\'{a}k, C.B.~Lang, T.~Neuhaus, and G.~Vones,
Phys. Lett. \textbf{155B}, 268 (1985).

\bibitem{Jansen:1985nh}
K.~Jansen, J.~Jers\'{a}k, C.B.~Lang, T.~Neuhaus, and G.~Vones,
Nucl. Phys. \textbf{B265}, 129 (1986).

\bibitem{Polley:1990tf}
L.~Polley and U.-J.~Wiese,
Nucl. Phys. \textbf{B356}, 629 (1991).

\bibitem{Evertz:1986nt}
H.G.~Evertz, V. Gr\"{o}sch, K.~Jansen, J.~Jers\'{a}k, H.A.~Kastrup, and T. Neuhaus,
Nucl. Phys. \textbf{B285}, 559 (1987).

\bibitem{Evertz:1986ur}
H.G.~Evertz, K.~Jansen, J.~Jers\'{a}k, C.B.~Lang, and T.~Neuhaus,
Nucl. Phys. \textbf{B285}, 590 (1987).

\bibitem{Hasenfratz:1988kr}
A.~Hasenfratz, K.~Jansen, J.~Jers\'{a}k, C.B.~Lang, T.~Neuhaus, and H.~Yoneyama,
Nucl. Phys. \textbf{B317}, 81 (1989).

\bibitem{Elitzur:1975im}
S.~Elitzur,
Phys. Rev. D \textbf{12}, 3978 (1975).


\end{thebibliography}
\end{document}